\documentclass[12pt,letterpaper]{article}
\usepackage{amsmath,amsfonts,amsthm,amssymb}

\swapnumbers 
\newtheorem{lemma}{Lemma}[section]
\newtheorem{corollary}[lemma]{Corollary}
\newtheorem{theorem}[lemma]{Theorem}
\newtheorem{definition}[lemma]{Definition}
\newtheorem{example}[lemma]{Example}

\newcommand{\Nat}{{\mathbb N}}
\newcommand\integers{{\mathbb Z}}

\newcommand\reals{{\mathbb R}}

\newcommand\hilb{{\mathfrak H}}

\begin{document}

\title{Rings with Effects}
\author{David J. Foulis{\footnote{Emeritus Professor of 
Mathematics and Statistics, University of Massachusetts, 
Postal Address: 1 Sutton Court, Amherst, MA 01002, USA; 
foulis@math.umass.edu.}}} 

\date{}
\maketitle

\begin{abstract}
An e-ring is a pair $(R,E)$ consisting of an associative ring $R$ 
with unity $1$ together with a subset $E\subseteq R$ of elements, 
called effects, with properties suggested by the so-called effect 
operators on a Hilbert space. We establish the basic properties of 
e-rings, investigate commutative e-rings called c-rings, relate 
certain c-rings called b-rings to Boolean algebras, and prove a 
structure theorem for b-rings.
\end{abstract}

\medskip


\medskip


\section{Introduction}  

Let $\hilb$ be a Hilbert space. In what follows, ${\mathbb B}(\hilb)$ 
denotes the $\sp{\ast}$-algebra of all bounded linear operators on 
$\hilb$, and ${\mathbb G}(\hilb)\subseteq{\mathbb B}(\hilb)$ is  
the subgroup of the additive group of ${\mathbb B}(\hilb)$ consisting 
of all Hermitian operators on $\hilb$.  The additive group 
${\mathbb G}(\hilb)$, organized into a partially ordered abelian group 
as usual, will be called the \emph{Hermitian group} for $\hilb$. The 
identity operator ${\mathbf 1}$ belongs to ${\mathbb G}(\hilb)$ and 
satisfies ${\mathbf 0}\leq{\mathbf 1}$.  We define ${\mathbb E}
(\hilb) :=\{E\in{\mathbb G}(\hilb)\mid{\mathbf 0}\leq E\leq{\mathbf 1}\}$ 
and (following G. Ludwig \cite{Lud}) we refer to operators in 
${\mathbb E}(\hilb)$ as \emph{effect operators} on $\hilb$. We also 
define ${\mathbb P}(\hilb) :=\{P\in{\mathbb G}(\hilb)\mid P\sp{2}=P\}$ 
to be the set of all (orthogonal) \emph{projection operators} on $\hilb$. 
Then we have
\[
{\mathbf 0},{\mathbf 1}\in{\mathbb P}(\hilb)\subseteq{\mathbb E}(\hilb)
 \subseteq{\mathbb G}(\hilb)\subseteq{\mathbb B}(\hilb).
\]

In mathematical physics, the representation of observables by 
so-called POV-measures, i.e., ${\mathbb E}(\hilb)$-valued measures on 
$\sigma$-fields \cite{BLM}, as well as by the more conventional 
${\mathbb P}(\hilb)$-valued measures, has now become a commonplace.  
Consequently, ${\mathbb E}(\hilb)$, ${\mathbb P}(\hilb)$, and 
suitable generalizations thereof, have come to be employed as 
algebraic models for the semantics of both sharp and unsharp 
quantum logics \cite{CMW,DGG,DP,FGB,GF,GPBB,LM}. Thus motivated, 
we take the pair $({\mathbb B}(\hilb),{\mathbb E}(\hilb))$ 
consisting of the ring ${\mathbb B}(\hilb)$ and the ``effect algebra" 
${\mathbb E}(\hilb)$ as a prototype for the following more general 
notion of an ``e-ring." 

\begin{definition} \label{def:eoring}
\textnormal{An \emph{e-ring} is a pair $(R,E)$ consisting of an 
associative ring $R$ with unity $1$ and a subset $E\subseteq R$ of 
elements called \emph{effects} such that $0,1\in E$; $e\in E
\Rightarrow 1-e\in E$; and the set $E\sp{+}$ consisting of all finite 
sums $e\sb{1}+e\sb{2}+\cdots+e\sb{n}$ with $e\sb{1},e\sb{2},...,e\sb{n}
\in E$ satisfies the following conditions: For all $a,b\in E\sp{+}$;}\\  

\begin{tabular}{ll}
{\rm (i)} $-a\in E\sp{+}\Rightarrow a=0$, & 
{\rm (ii)} \ $1-a\in E\sp{+}\Rightarrow a\in E$,\\ 
{\rm (iii)} $ab=ba\Rightarrow ab\in E\sp{+}$, & 
{\rm (iv)} $aba\in E\sp{+}$,\\ 
{\rm (v)} \ $aba=0\Rightarrow ab=ba=0$, and &  
{\rm (vi)}  $(a-b)\sp{2}\in E\sp{+}$.
\end{tabular}\\ 

\end{definition}

The notion of an e-ring is mathematically equivalent to the notion 
of an effect-ordered ring originally introduced in \cite{FCPOAG}, but 
reformulated to emphasize the role of the ``effect algebra" $E$.

It is easy to see that the prototypic pair $({\mathbb B}(\hilb),
{\mathbb E}(\hilb))$ is an e-ring as soon as one observes that 
${\mathbb E}(\hilb)\sp{+}$ is precisely the set of all positive 
Hermitian operators on $\hilb$. (By a slight abuse of language, we 
call a Hermitian operator $A$ ``positive" if $A\geq{\mathbf 0}$.) 
Indeed, every effect operator is positive by definition, and a finite 
sum of positive Hermitian operators is positive. Conversely, if $A$ 
is a positive Hermitian operator, $\|A\|$ is the uniform operator 
norm of $A$, and $n$ is a positive integer with $n\geq\|A\|$, then 
$(1/n)A$ is an effect operator, and $A$ is a sum of $n$ effect 
operators, each equal to $(1/n)A$.

In addition to the prototypic example $({\mathbb B}(\hilb),
{\mathbb E}(\hilb))$, we have the following simple examples 
of e-rings. (More examples will be given later.) We denote the 
set of positive integers by $\Nat :=\{1,2,3,...\}$.  

\begin{example} \label{ex:Eis01}
Let $R$ be a ring with unity $1$ such that $n\cdot 1\not=0$ for all 
$n\in\Nat$, and define $E :=\{0,1\}$. Then $(R,E)$ is an e-ring.
\end{example}

\begin{example} \label{ex:subfield}
If $R$ is any subfield of the totally ordered field $\reals$ of real 
numbers and $E :=\{e\in R\mid 0\leq e\leq 1\}$, then $(R,E)$ is an 
e-ring.
\end{example}

As Examples \ref{ex:Eis01} and \ref{ex:subfield} illustrate, 
the definition of an e-ring $(R,E)$ does not necessarily provide 
a strong connection between the algebraic structure of the set 
$E$ of effects and the ring structure of $R$.  For instance, if 
$(R,E)$ is an e-ring and ${\widetilde R}$ is any extension ring 
of $R$ such that $1\cdot x=x\cdot 1=x$ for all $x\in{\widetilde R}$, 
then $({\widetilde R},E)$ is also an e-ring. The rather weak 
connection between $E$ and $R$ does not concern us here because, 
motivated by quantum measurement theory and quantum logic, we are 
mainly interested in the structure of the ``effect algebra" $E$, 
and the enveloping ring $R$ is just a convenient environment 
in which to study this structure.

Section 2 below, which treats the basic properties of e-rings, 
culminates with a theorem that the set $P$ of projections in an e-ring 
acquires (at least) the structure of an orthomodular poset 
(Theorem \ref{th:OMP}) and a theorem stating that $P$ indexes a 
so-called compression base for the directed group generated by the 
effects (Theorem \ref{th:compbase}). Section 3, which treats 
commutativity and coexistence in the effect algebra $E$ of an e-ring, 
culminates in a structure theorem for a class of e-rings (called 
b-rings) in which every effect is a projection (Theorem 
\ref{th:bring}). In a subsequent paper, we shall study square 
roots and polar decompositions in e-rings.

\section{Basic Properties of e-Rings} 

We begin by extracting from an e-ring $(R,E)$ an analogue $G$ 
of the Hermitian group ${\mathbb G}(\hilb)$ for the prototype
$({\mathbb B}(\hilb),{\mathbb E}(\hilb))$.

\begin{theorem} \label{th:G}
Let $(R,E)$ be an e-ring and let $E\sp{+}$ be the set of all 
sums of finite sequences f elements of $E$. Then
\[ 
G :=E\sp{+}-E\sp{+}=\{a-b\mid a,b\in E\sp{+}\}
\]
is a subgroup of the additive group of the ring $R$, and $G$ 
is a directed partially ordered abelian group with positive 
cone $E\sp{+}=\{g\in G\mid 0\leq g\}$. Moreover, $E\subseteq G$ 
and $E$ generates the group $G$.
\end{theorem}

\begin{proof}
Since $E\sp{+}$ is closed under addition, it is clear that 
$G=E\sp{+}-E\sp{+}$ is a subgroup of the additive group of 
the ring $R$.  Also, $0\in E\subseteq E\sp{+}\subseteq G$, 
and by Definition \ref{def:eoring} (i), if both $a$ and $-a$ 
belong to $E\sp{+}$, then $a=0$. Therefore, $G$ is a partially 
ordered abelian group with positive cone $E\sp{+}$, the partial 
order being given by $g\leq h\Leftrightarrow h-g\in E\sp{+}$ for 
$g,h\in G$ \cite[p. 3]{Good}. By the definition of $G$, every 
element $g\in G$ can be written in the form $g=a-b$ with $a,b
\in E\sp{+}$, i.e., $G$ is directed \cite[p. 4]{Good}. Thus, 
$E\sp{+}$ generates the group $G$, and since $E$ generates 
$E\sp{+}$ as an additive semigroup, it follows that $E$ 
generates $G$ as a group.
\end{proof}

If $(R,E)$ is an e-ring then, \emph{as an additive abelian group}, 
$R$ is partially ordered (but not necessarily directed) with 
$E\sp{+}$ as its positive cone; however, unless $R$ is commutative, 
it is not necessarily a partially ordered ring (as usually 
understood) because $E\sp{+}$ need not be closed under multiplication.

\begin{definition} \label{def:GandP}
\textnormal{Let $(R,E)$ be an e-ring and let $E\sp{+}$ be the set 
of all sums of finite sequences of elements of $E$.  Then:}  
\begin{enumerate}

\item\textnormal{The partially ordered additive abelian group $G :=
 E\sp{+}-E\sp{+}$ (Theorem \ref{th:G}) is called the \emph{directed 
group} of $(R,E)$.}
\item\textnormal{Idempotent elements $p=p\sp{2}\in G$ are called 
 \emph{projections}.}  

\end{enumerate}
\end{definition}

For the prototype e-ring $({\mathbb B}(\hilb),{\mathbb E}(\hilb))$,   
the directed group is the Hermitian group ${\mathbb G}(\hilb)$, and 
${\mathbb P}(\hilb)$ is the set of projections. In Example 
\ref{ex:Eis01}, the directed group $G=\{n\cdot 1\mid n\in\integers\}$ 
of $(R,E)$ is isomorphic to the totally ordered additive group of the 
ring $\integers$ of integers. In Example \ref{ex:subfield}, $G$ is 
the additive subgroup of the field $R$ with the total order inherited 
from $\reals$. In both Examples \ref{ex:Eis01} and \ref{ex:subfield}, 
the only projections are $0$ and $1$.

The e-ring $({\mathbb B}(\hilb),{\mathbb E}(\hilb))$ is a special 
case (when $A$ is a type-I von Neumann factor) of the e-ring $(A,E)$ 
in the following example.

\begin{example} \label{ex:CStar}
Let $A$ be a unital C\,$\sp{\ast}$-algebra and let 
\[
E :=\{aa\sp{\ast}\mid a\in A\text{ and }\exists\,b\in A, aa\sp{\ast}
 +bb\sp{\ast}=1\}.
\]
Then $(A,E)$ is an e-ring, $E\sp{+}=\{aa\sp{\ast}\mid a\in A\}$, 
and the directed group $G$ for $(A,E)$ is the additive group of 
self-adjoint elements in $A$.
\end{example}

\noindent{\bf In the sequel, we assume that $(R,E)$ is an e-ring, 
$1$ is the unity element in $R$, $E\sp{+}$ is the set of all sums 
of finite sequences of elements of $E$, $G=E\sp{+}-E\sp{+}$ is 
the directed group of $(R,E)$, $G$ is partially ordered with 
positive cone $E\sp{+}$, and
\[
P=\{p\in G\mid p=p\sp{2}\}
\]
is the set of all projections in $G$. It is understood that $P$ and 
$E$ are partially ordered by the restrictions of the partial order 
$\leq$ on $G$.}

\begin{lemma} \label{lm:AA} Let $g,h\in G$ and let $p\in P$. 
Then: 
\smallskip

\begin{tabular}{lll}
{\rm (i)} \ \ $g\sp{2}\in E\sp{+}$.  & 
{\rm (ii)} \  $gh+hg\in G$. &
{\rm (iii)} $g\in E\sp{+}\Rightarrow ghg\in G$. \\ 
{\rm (iv)} $php\in G$. &
{\rm (v)} \  $h\in E\sp{+}\Rightarrow php\in E\sp{+}$. 
\end{tabular}

\end{lemma}

\begin{proof} (i) By hypothesis, $G$ is directed, hence $g$ 
is a difference of two elements in $E\sp{+}$, so $g\sp{2}\in 
E\sp{+}$ by Definition \ref{def:eoring} (vi).

(ii) By (i), $gh+hg=(g+h)\sp{2}-g\sp{2}-h\sp{2}\in G$.

(iii) As $G$ is directed, there exist $a,b\in E\sp{+}$ such 
that $h=a-b$. By Definition \ref{def:eoring} (iv), $gag,gbg
\in E\sp{+}$, whence $ghg=gag-gbg\in G$. 

(iv) As $p=p\sp{2}\in G$, (i) implies that $p\in E\sp{+}$; 
 hence (iv) follows from (iii).

(v) As in (iv), $p\in E\sp{+}$, so (v) follows from Definition 
\ref{def:eoring} (iv).
\end{proof}

\begin{lemma} \label{lm:E}
{\rm (i)} $E=\{e\in G\mid 0\leq e\leq 1\}$. {\rm (ii)} $0,1\in P$. {\rm (iii)} 
$p\in P\Rightarrow 1-p\in P$. {\rm (iii)} $P\subseteq E$.
\end{lemma}

\begin{proof}
(i) By Definition \ref{def:eoring}, we have $0,1\in E$. Evidently, 
if $e\in E$, then $e,1-e\in E\subseteq E\sp{+}=\{g\in G\mid 0\leq g\}$, 
whence $e\in G$ with $0\leq e\leq 1$. Conversely, suppose $e\in G$ 
with $0\leq e\leq 1$. Then $e,1-e\in E\sp{+}$, and it follows from 
Definition \ref{def:eoring} (ii) that $e\in E$. Thus, $E=\{e\in G
\mid 0\leq e\leq 1\}$.

(ii) $0=0\sp{2}\in G$ and $1=1\sp{2}\in G$.

(iii) If $p\in G$ with $p=p\sp{2}$, then $1-p\in G$ with 
$(1-p)\sp{2}=1-2p+p\sp{2}=1-p$.

(iv) Suppose $p\in P$. By Lemma \ref{lm:AA} (i), $p=p\sp{2}\in 
E\sp{+}$, i.e., $0\leq p$. Thus, by (iii), $0\leq 1-p$, whence 
$0\leq p\leq 1$, so $p\in E$ by (i).   
\end{proof}

In view of Lemma \ref{lm:E} (i), we shall refer to $E$ as the 
\emph{unit interval} in $G$. We have 
\[
0,1\in P\subseteq E\subseteq E\sp{+}\subseteq G\subseteq R.
\]
Equipped with the partially defined binary operation $\oplus$ 
obtained by restricting $+$ on $G$ to $E$, the unit interval $E$ 
forms a so-called \emph{effect algebra} \cite{BF97}.  The 
effect algebras arising from e-rings in this way are rather special
in that they admit a (perhaps only partial) multiplicative structure 
(cf. \cite{D02}).

\begin{lemma} \label{lm:FF}
Let $d,e,f\in E$ with $ef=fe$. Then:\smallskip\\
\begin{tabular}{ll}
{\rm (i)} \ \ $0\leq ef\leq e,f\leq 1$. & 
{\rm (ii)} $0\leq ede\leq e\sp{2}\leq e\leq 1$. \smallskip\\
{\rm (iii)} $0\leq e,f\leq e+f-ef\leq 1$. & 
{\rm (iv)} $0\leq e-e\sp{2}\leq e,1-e\leq 1$.
\end{tabular}
\end{lemma}

\begin{proof} Assume the hypotheses. 

(i) By Definition \ref{def:eoring} (iii), $0\leq ef$. Likewise, 
$0\leq e(1-f)=e-ef$, so $ef\leq e$, and by symmetry, 
$ef\leq f$.

(ii) By Definition \ref{def:eoring} (iv), $0\leq ede$ and 
$0\leq e(1-d)e=e\sp{2}-ede$.  Also, by (i) with $f=e$, we have 
$e\sp{2}\leq e\leq 1$.

(iii) By (i), $0\leq f-ef$, so $e\leq e+f-ef$, and by symmetry, 
$f\leq e+f-ef$. Also, by Definition \ref{def:eoring} (iii), 
$0\leq(1-e)(1-f)=1-e-f+ef$, whence $e+f-ef\leq 1$.

(iv) By (ii), $0\leq e-e\sp{2}$. Also, by Lemma \ref{lm:AA} 
(i), $0\leq e\sp{2}$, whence $e-e\sp{2}\leq e$. Finally, by (iii) 
with $f=e$, we have $2e-e\sp{2}\leq 1$, so $e-e\sp{2}\leq 1-e$.
\end{proof}

\begin{lemma} \label{lm:M}
Let $g,h,k\in E\sp{+}$, $p\in P$, and $n\in\Nat$. Then: 
{\rm (i)} $gh=0\Rightarrow hg=0$. {\rm (ii)} If 
$gk=kg$ and $hk=kh$, then $g\leq h\Rightarrow gk\leq hk$. 
{\rm (iii)} If $gh=hg$, then $g\leq h\Rightarrow g\sp{2}
\leq h\sp{2}$. {\rm (iv)} $g\leq np\Rightarrow g=gp=pg$. 
{\rm (v)} $g\sp{n}=0\Rightarrow g=0$.
\end{lemma}

\begin{proof}
(i) $gh=0\Rightarrow ghg=0\Rightarrow hg=0$ by Definition 
\ref{def:eoring} (v).

(ii) Assume the hypotheses. Then $h-g\in E\sp{+}$ and 
$hk-gk=(h-g)k=k(h-g)\in E\sp{+}$ by Definition \ref{def:eoring} 
(iii).

(iii) Assume the hypotheses. By (ii), $g\sp{2}\leq gh$ and 
$gh\leq h\sp{2}$, so $g\sp{2}\leq h\sp{2}$.

(iv) Assume the hypotheses. Then $g,np-g\in E\sp{+}$, whence 
$(1-p)g(1-p),(1-p)(np-g)(1-p)=-(1-p)g(1-p)\in E\sp{+}$ by 
Lemma \ref{lm:AA} (v).  Therefore, $(1-p)g(1-p)=0$ by 
Definition \ref{def:eoring} (i), and it follows from Definition 
\ref{def:eoring} (v) that $(1-p)g=g(1-p)=0$, i.e., $g=pg=gp$. 

(v) We may assume that $n$ is the smallest positive integer such 
that $g\sp{n}=0$. If $n$ is even and $k=n/2$, we have $g\sp{k}
\cdot 1\cdot g\sp{k}=0$, so $g\sp{k}=g\sp{k}\cdot 1=0$ by 
Definition \ref{def:eoring} (v), contradicting our assumption.
Therefore $n$ is odd. If $n=1$, we are done, so we may assume that 
$n=2k+1$ where $k\in\Nat$. Then $g\sp{k}gg\sp{k}=0$, so $g\sp{k+1}
=g\sp{k}g=0$ by Definition \ref{def:eoring} (v), again contradicting 
our assumption. 
\end{proof}

According to part (i) of the following lemma, $1$ is a so-called 
\emph{order unit} in $G$ \cite[p. 4]{Good}.

\begin{lemma} \label{lm:orderunit}
{\rm (i)} If $g\in G$, there exists $n\in\Nat$ such that $g\leq n
\cdot 1$. {\rm (ii)} If $a\sb{1},a\sb{2},...,a\sb{n}\in E\sp{+}$ 
and $a\sb{1}+ a\sb{2}+\cdots+a\sb{n}=0$, then $a\sb{1}=a\sb{2}=\cdots
=a\sb{n}=0$.
\end{lemma}

\begin{proof}
(i) Write $g=a-b$ with $0\leq a,b$. Then $0\leq b=a-g$, whence 
$g\leq a$. As $a\in E\sp{+}$, there exist $e\sb{1},e\sb{2},
...,e\sb{n}\in E$ with $a=e\sb{1}+e\sb{2}+\cdots+e\sb{n}$. 
By Lemma \ref{lm:E} (i), $e\sb{i}\leq 1$ for $i=1,2,...,n$, 
and it follows that $g\leq a\leq n\cdot 1$.

(ii) Assume the hypotheses. It will be sufficient to prove that 
$a\sb{1}=0$. But, $-a\sb{1}=a\sb{2}+\cdots+a\sb{n}\in E\sp{+}$, 
so $a\sb{1}=0$ by Definition \ref{def:eoring} (i).
\end{proof}

\begin{theorem} \label{th:CC}
Let $e\in E$ and $p\in P$. Then the following conditions are 
mutually equivalent: {\rm (i)} $e\leq p$, {\rm (ii)} $e=ep=pe$, 
{\rm (iii)} $e=pep$, {\rm (iv)} $e=ep$, {\rm (v)} $e=pe$.
\end{theorem}

\begin{proof}
(i) $\Rightarrow$ (ii). Assume that $e\leq p$ and let $d :=p-e$. 
Then $e,d\in E\sp{+}$, $e+d=p$, and 
\[
(1-p)e(1-p)+(1-p)d(1-p)=(1-p)p(1-p)=0.
\]
By Lemma \ref{lm:E} (iii), $1-p\in P$, whence by Lemma \ref{lm:AA} 
(v), 
\[
(1-p)e(1-p),(1-p)d(1-p)\in E\sp{+},
\]
and it follows from Lemma \ref{lm:orderunit} (ii) that $(1-p)e(1-p)=(1-p)d(1-p)
=0$. Therefore, by Definition \ref{def:eoring} (v), $(1-p)e=e(1-p)=0$, 
i.e., $e=pe=ep$. 

(ii) $\Rightarrow$ (iii) $\Rightarrow$ (iv). Follows from $p=p\sp{2}$.

(iv) $\Leftrightarrow$ (v). By Lemma \ref{lm:M} (i), 
$e=pe\Rightarrow (1-p)e=0\Rightarrow e(1-p)=0\Rightarrow e=ep$, 
and the converse implication follows by symmetry.

(v) $\Rightarrow$ (i).  Assume (v). Since (iv) 
$\Leftrightarrow$ (v), we have $pe=ep=e$, so $(1-e)p=p(1-e)
=p-e$, whence  $p-e\in E\sp{+}$ by Definition \ref{def:eoring} 
(iii), and therefore $e\leq p$.
\end{proof}

\begin{corollary} \label{cor:DD}
Let $e\in E$ and $p\in P$. Then the following conditions are 
mutually equivalent: {\rm (i)} $p\leq e$, {\rm (ii)} $p=ep=pe$, 
{\rm (iii)} $p+pep=pe+ep$, {\rm (iv)} $p=ep$, {\rm (v)} $p=pe$.
\end{corollary}

\begin{proof}
Replace $e$ by $1-e$ and $p$ by $1-p$ in Theorem \ref{th:CC}, noting 
that $p\leq e\Leftrightarrow 1-e\leq 1-p$.
\end{proof}

\begin{theorem} \label{th:p+qinP}
Let $p,q\in P$.  Then the following conditions are mutually 
equivalent: {\rm (i)} $p+q\in E$, {\rm (ii)} $p+q\leq 1$, 
{\rm (iii)} $pq=0$, {\rm (iv)} $pq=qp=0$, {\rm (v)} 
$p+q\in P$.
\end{theorem}

\begin{proof}
(i) $\Rightarrow$ (ii) Follows from Lemma \ref{lm:E} (i).

(ii) $\Rightarrow$ (iii). If $p+q\leq 1$, then $p\leq 1-q$, 
and it follows from Theorem \ref{th:CC} that $p=p(1-q)=p-pq$, 
whence $pq=0$.

(iii) $\Rightarrow$ (iv). Follows from Lemma \ref{lm:M} (i).

(iv) $\Rightarrow$ (v). If $pq=qp=0$, then $(p+q)\sp{2}=
p\sp{2}+pq+qp+q\sp{2}=p+q$.

(v) $\Rightarrow$ (i). Follows from Lemma \ref{lm:E} (iv).
\end{proof}

\begin{theorem} \label{th:pqinP}
If $p,q\in P$, then the following conditions are 
mutually equivalent: {\rm (i)} $pq\in P$, {\rm (ii)} $pq\in 
E$, {\rm (iii)} $pq=qp$, {\rm (iv)} $pq=pqp$.
Moreover, if any---hence all---of these conditions hold, 
then $pq=p\wedge q$ is the infimum {\rm (}greatest lower 
bound{\rm )} of $p$ and $q$ both in $P$ and in $E$.
\end{theorem}

\begin{proof}
(i) $\Rightarrow$ (ii) is obvious.

(ii) $\Rightarrow$ (iii). Assume that $e :=pq\in E$. Since 
$e=pe=eq$, Theorem \ref{th:CC} implies that $e\leq p,q$  
and $e=ep=qe$.  Thus, $e\sp{2}=epq=e$, so $e\in P$. As 
$e\leq p,q$, it follows that $p-e,q-e\in E\sp{+}$.  
Furthermore, $(p-e)(q-e)=pq-pe-eq+e\sp{2}=0$,
whence Lemma \ref{lm:M} (i) implies that 
$0=(q-e)(p-e)=qp-e$, i.e., $qp=e=pq$. 

(iii) $\Rightarrow$ (iv). If $pq=qp$, then $pq=p(pq)=pqp$.

(iv) $\Rightarrow$ (i). If $pq=pqp$, then $(pq)\sp{2}=
pqpq=(pq)q=pq$, so $pq\in P$.

Suppose that any, hence all of the conditions (i)--(iv) hold.
Then $pq\in P$ with $pq\leq p,q$. Also, if $e\in E$ with 
$e\leq p,q$, then $e=ep=eq$, whence $e(pq)=e$, i.e., 
$e\leq pq$.
\end{proof}

\begin{corollary} \label{cor:pqinP}
If $p,q\in P$ and $pq=qp$, then $p+q-pq\in P$ and 
$p+q-pq=p\vee q$ is the supremum {\rm (}least upper bound{\rm )} 
of $p$ and $q$ both in $P$ and in $E$.
\end{corollary}

\begin{proof}
The mapping $r\mapsto(1-r)$ is order inverting and of period 
$2$ on $P$. Also $pq=qp\Leftrightarrow(1-p)(1-q)=(1-q)(1-p)$,  
and $1-(1-p)(1-q)=p+q-pq$.
\end{proof}

\begin{corollary} \label{cor:q-pinP}
Let $p,q\in P$. Then $q-p\in E\Leftrightarrow p\leq q 
\Leftrightarrow q-p\in P$. Moreover, if $p\leq q$, then 
$q-p=q\wedge(1-p)$ is the infimum of $q$ and $1-p$ both 
in $P$ and in $E$.
\end{corollary}

\begin{proof}
If $q-p\in E$, then $0\leq q-p$, whence $p\leq q$. 
Suppose that $p\leq q$. Then $p=pq=qp$ by Theorem \ref{th:CC}, 
so $q(1-p)=(1-p)q=q-p$ and it follows from Theorem \ref{th:pqinP}  
that $q-p\in P$ and that $q-p$ is the infimum both in $E$ 
and $P$ of $q$ and $1-p$.
\end{proof}

\begin{theorem} \label{th:OMP}
With $p\mapsto 1-p$ as orthocomplementation, $P$ is an 
orthomodular poset {\rm (OMP) \cite{DGG,DP}} and, for $p,q\in P$ 
with $p\leq 1-q$, the supremum in $P$ of $p$ and $q$ 
is $p\vee q=p+q$.
\end{theorem}

\begin{proof}
We have $0\leq p\leq 1$ for all $p\in P$, and $p\mapsto 1-p$ is 
an order-reversing mapping of period $2$ on $P$. Let $p,q\in P$.  
If $p\leq 1-q$, then $p+q\leq 1$, whence $p+q\in E$, so 
$p+q\in P$ with $pq=qp=0$ by Theorem \ref{th:p+qinP}, and it 
follows from Corollary \ref{cor:pqinP} that $p+q=p+q-pq$ is the 
supremum of $p$ and $q$ in $P$. Now suppose that $p\leq q$. 
By Corollary \ref{cor:q-pinP}, $q-p\in P$, whence $q=p+(q-p)$ 
is the orthomodular identity.
\end{proof}

\begin{lemma} \label{lm:Pnormal}
Suppose that $d,e,f,d+e+f\in E$ with $d+e,d+f\in P$. Then $d,e,f\in P$.
\end{lemma}

\begin{proof}
Assume the hypotheses and let $p :=d+e\in P$ and $q :=d+f\in P$.
Then $p+f=d+e+f\leq 1$, so $f\leq 1-p\in P$; hence by Theorem 
\ref{th:CC}, $f=(1-p)f=f-pf$, and therefore $pf=0$. Also, $d\leq 
d+e=p$, so $pd=d$ by Theorem \ref{th:CC}. Consequently, $pq=
p(d+f)=d\in E$, and it follows from Theorem \ref{th:pqinP} that 
$d=pq=pq\in P$. As $d\leq p$ and $d,p\in P$, it follows from 
Corollary \ref{cor:q-pinP} that $e=p-d\in P$, and likewise $f=
q-d\in P$.
\end{proof}

The following theorem provides useful conditions---not directly 
involving multiplication---for determining whether an effect 
is a projection.

\begin{theorem} \label{th:sharp}
If $e\in E$, then the following conditions are mutually equivalent:
{\rm (i)} If $a,b,a+b\in E$, then $a,b\leq e\Rightarrow a+b\leq e$.
{\rm (ii)} If $d\in E$ with $d\leq e,1-e$, then $d=0$. {\rm (iii)}  
$e\in P$. 
\end{theorem}

\begin{proof}
(i) $\Rightarrow$ (ii). Assume (i) and the hypotheses of (ii).
Then $d+e\leq 1$ with $d,e\leq e$, whence $d+e\leq e$, and it 
follows that $d=0$.

(ii) $\Rightarrow$ (iii). Assume (ii).  By Lemma \ref{lm:FF} (iv), 
$0\leq e-e\sp{2}\leq e,1-e,$ so $e-e\sp{2}=0$, i.e., $e=e\sp{2}$.

(iii) $\Rightarrow$ (i).  Suppose $e\in P$ and assume the 
hypotheses of (i). By Corollary \ref{cor:DD}, $a=ae$ and $b=be$, 
whence $(a+b)e=a+b$, and it follows that $a+b\leq e$.
\end{proof}

If $p\in P$ and $g\in G$, then by Lemma \ref{lm:AA} (iv), $pgp\in G$; 
hence we can define the mapping $J\sb{p}\colon G\to G$ by 
$J\sb{p}(g)=pgp$ for all $g\in G$.  Thus, owing to Lemmas \ref{lm:AA}, 
\ref{lm:E}, \ref{lm:Pnormal}, and Theorem \ref{th:p+qinP}, we have 
the following theorem (see \cite{FDenver,FCB}). 

\begin{theorem} \label{th:compbase}
The family $(J\sb{p})\sb{p\in P}$ is a compression base for $G$.
\end{theorem}

The partially ordered abelian group $G$ is said to be 
\emph{archimedean} iff, whenever $g,h\in G$ and $ng\leq h$ 
for all $n\in\Nat$, it follows that $g\leq 0$ \cite[p. 20]{Good}. 
An order-preserving group endomorphism $J\colon G\to G$ is 
called a \emph{retraction} iff $J(1)\in E$ and, for all $e\in E$, 
$e\leq J(1)\Rightarrow J(e)=e$. If $p\in P$, it is clear that 
$J\sb{p}$ is a retraction on $G$. Conversely, as a consequence 
of \cite[Corollary 4.6]{FCPOAG}, we have the following theorem.

\begin{theorem}
If $G$ is archimedean, then every retraction $J$ on $G$ has the 
form $J=J\sb{p}$ with $p=J(1)\in P$.
\end{theorem}  

\section{Commuting Elements of $G$} 

\emph{We maintain our standing hypothesis that $(R,E)$ is an e-ring,
$G$ is its directed group, and $P$ is the OMP of projections 
in $G$.}

\begin{definition}
\textnormal{Let $g,h\in G$.  We write $gCh$ iff $gh=hg$ and we define 
the \emph{commutant} of $g$ in $G$ by $C(g) :=\{h\in G\mid gCh\}$. 
More generally, if $X\subseteq G$, then $C(X) :=\bigcap\sb{x\in X}C(x)$ 
is called the \emph{commutant} of $X$.}
\end{definition}

In contrast with more-or-less standard usage, e.g., in operator theory, 
we use the notion of the commutant \emph{only in relation to 
elements of} $G$, and not to general elements of the enveloping 
ring $R$.

If $L$ is any OMP, then two elements $p,q\in L$ are said to be 
\emph{Mackey compatible} iff there exist pairwise orthogonal 
elements $p\sb{1},q\sb{1},d\in L$ with $p=d\vee p\sb{1}$, and 
$q=d\vee q\sb{1}$ \cite{DP}. By Theorem \ref{th:OMP}, projections $p,q$ 
in the OMP $P$ are Mackey compatible iff there exist projections 
$p\sb{1},q\sb{1},d\in P$ with $d+p\sb{1}+q\sb{1}\leq 1$, 
$p=d+p\sb{1}$, and $q=d+q\sb{1}$. The next lemma provides a useful 
condition---not directly involving multiplication---for 
determining whether two projections commute.

\begin{lemma} \label{lm:pCq}
If $p,q\in P$, then $pCq$ iff $p$ and $q$ are Mackey compatible in 
$P$.
\end{lemma} 

\begin{proof}
Suppose that $pCq$. By Theorem \ref{th:pqinP}, $pq\in P$ with 
$pq\leq p,q$; by Corollary \ref{cor:pqinP}, $p+q-pq\in P$; 
and by Corollary \ref{cor:q-pinP}, $p\sb{1} :=p-pq\in P$ and 
$q\sb{1} :=q-pq\in P$.  Thus, with $d :=pq$, we have $d+
p\sb{1}+q\sb{1}=p+q-pq\in P$ with $p=d+p\sb{1}$, and $q=d+q\sb{1}$.

Conversely, suppose there exist $p\sb{1},q\sb{1},d\in P$ 
such that $d+p\sb{1}+q\sb{1}\in P$, $p=d+p\sb{1}$, and 
$q=d+q\sb{1}$. Then $d+p\sb{1}\leq d+p\sb{1}+q\sb{1}\leq 1$, 
whence $dCp\sb{1}$ by Theorem \ref{th:p+qinP}. Likewise, 
$dCq\sb{1}$, and $p\sb{1}Cq\sb{1}$, whence $pCq$.
\end{proof}

In the following definition, the condition in Lemma \ref{lm:pCq} 
is generalized to effects $e,f\in E$.

\begin{definition}
\textnormal{Effects $e,f\in E$ are said to be \emph{coexistent} iff 
there exist effects $d,e\sb{1},f\sb{1}\in E$ such that $d+e\sb{1}+f\sb{1}
\in E$, $e=d+e\sb{1}$, and $f=d+f\sb{1}$.}
\end{definition}

The terminology ``coexistent" is borrowed from the quantum theory of 
measurement \cite{BLM}. (Some authors also refer to coexistent effects 
as being ``Mackey compatible," but, since coexistent effects need not 
commute, we prefer not to follow this practice.)

\begin{lemma} \label{lm:CimpliesCE}
Let $e,f\in E$. Then: {\rm (i)} If $eCf$, then $e$ and $f$ are 
coexistent. {\rm (ii)} If $e+f\leq 1$, then $e$ and $f$ are coexistent.
\end{lemma}

\begin{proof}
(i) Let $d :=ef$, $e\sb{1} :=e-ef$, and $f\sb{1} :=f-ef$. By Lemma 
\ref{lm:FF} (i), $d,e\sb{1},f\sb{1}\in E$. Also, $d+e\sb{1}+f\sb{1}=
e+f-ef\in E$ by Lemma \ref{lm:FF} (iii).

(ii) If $e + f\in E$, then $0+e+f\in E$ with $e=0+e$ and $f=0+f$.
\end{proof}

In general, the converse of Lemma \ref{lm:CimpliesCE} (i) is false. 
For instance, in the prototype ${\mathbb E}(\hilb)$, choose two 
effect operators $A$ and $B$ that do not commute. Then $\frac{1}{2}A$ 
and $\frac{1}{2}B$ are non-commuting effect operators; yet, since 
$\frac{1}{2}A+\frac{1}{2}B\leq{\mathbf 1}$, they are coexistent.
However, we do have the following result.

\begin{theorem} 
Let $p,q\in P$. Then, regarded as effects in $E$, the projections 
$p$ and $q$ are coexistent iff $pCq$.
\end{theorem} 

\begin{proof}
Combine Lemma \ref{lm:Pnormal} and Lemma \ref{lm:pCq}.
\end{proof}

In a Boolean algebra (i.e., a bounded complemented distributive 
lattice), every element has a unique complement; hence 
if an OMP is a Boolean algebra, the Boolean complementation    
mapping coincides with the orthocomplementation mapping. 
It is well known that an OMP is a Boolean algebra iff its 
elements are pairwise Mackey compatible \cite{DP}; hence 
we have the following.

\begin{corollary} \label{cor:PBoo}
The OMP $P$ is a Boolean algebra iff $P\subseteq C(P)$. 
Moreover, if $P$ is a Boolean algebra, then $p\mapsto 
1-p$ is the Boolean complementation mapping on $P$. 
\end{corollary}

In what follows, we shall be considering the condition 
$G\subseteq C(G)$ and the weaker condition $G\subseteq C(P)$. 
For instance, if the enveloping ring $R$ is commutative, then 
$G\subseteq C(G)$. Since $E$ generates the group $G$, it follows 
that $G\subseteq C(G)\Leftrightarrow E\subseteq C(E)$ and that 
$G\subseteq C(P)\Leftrightarrow E\subseteq C(P)$. Also, if 
$G\subseteq C(P)$, then $P\subseteq C(P)$, whence $P$ is a 
Boolean algebra by Corollary \ref{cor:PBoo}. If the unital 
C$\sp{\ast}$-algebra $A$ in Example \ref{ex:CStar} satisfies 
virtually any version of the spectral theorem (e.g., if $A$ is a 
von Neumann algebra, or even an AW$\sp{\ast}$-algebra), then 
$P\subseteq C(P)$ will imply that $G\subseteq C(G)$.

Suppose that $G\subseteq C(G)$, let $g,h\in G$, and choose 
$a,b,c,d\in E\sp{+}$ such that $g=a-b$ and $h=c-d$. By Definition 
\ref{def:eoring} (iii), $ac,ad,bc,bd\in E\sp{+}\subseteq G$, 
whence $gh=ac-ad-bc+bd\in G$, and it follows that $G$ is not only an 
additive abelian group, but a commutative subring of $R$. Clearly,  
with $G$ thus organized into a ring, $(G,E)$ is an e-ring, $G$ is a 
partially ordered commutative ring with unity $1$, the partially 
ordered additive group $G$ is the directed group of $(G,E)$, 
and $(G,E)$ is a \emph{c-ring} as per the following definition.

\begin{definition}  \label{def:cring}
\textnormal{A \emph{c-ring} is an e-ring $(G,E)$ such that $G$ is 
a commutative ring and $G=E\sp{+}-E\sp{+}$.}
\end{definition}

If $G\subseteq C(G)$ and $R\not=G$, \emph{we can disregard the 
enveloping ring $R$ and drop down to the c-ring $(G,E)$.} Evidently, 
the passage from the e-ring $(R,E)$ to the c-ring $(G,E)$ affects 
neither the structure of the effect algebra $E$ nor of the Boolean 
algebra $P$.

As a consequence of the Gelfand representation theorem \cite
[Theorem 4.4.3]{KR}, the following example of a c-ring may be 
regarded as the commutative version of Example \ref{ex:CStar}. 

\begin{example} \label{ex:CofX}
Let $X$ be a compact Hausdorff space, define $C(X,\reals)$ to be 
the ring of all continuous real-valued functions $f\colon X\to\reals$ 
with pointwise operations, and let 
\[ 
E(X,\reals) :=\{e\in C(X,\reals)\mid 0\leq e(x)\leq 1, \forall x
 \in X\}.
\] Then $(C(X,\reals),E(X,\reals))$ is a c-ring, the partial order 
on $C(X,\reals)$ is the pointwise partial order, $C(X,\reals)$ is 
archimedean, and the Boolean algebra 
\[
P(X,\reals) :=\{p\in C(X,\reals)\mid p(x)\in
\{0,1\},\forall x\in X\}
\] 
of projections in $(C(X,\reals),E(X,\reals))$ consists of the 
characteristic set functions $\chi\sb{K}$ of compact open subsets 
$K$ of $X$.
\end{example}

In the following example of a c-ring, the effects are ``fuzzy 
subsets" of $X$ in the sense of L. Zadeh \cite{Zad}.

\begin{example} \label{ex:sigmafield}
Let ${\cal F}$ be a $\sigma$-field of subsets of a nonempty 
set $X$, define ${\cal B}(X,{\cal F},\reals)$ to be the 
ring under pointwise operations of all bounded real-valued  
${\cal F}$-measurable functions $f\colon X\to\reals$, and let 
\[ 
{\cal E}(X,{\cal F},\reals) :=\{e\in{\cal B}(X,{\cal F},
 \reals)\mid 0\leq e(x)\leq 1,\forall x\in X\}.
\] Then $({\cal B}(X,{\cal F},\reals),{\cal E}(X,{\cal F},
\reals))$ is a c-ring, the partial order on ${\cal B}
(X,{\cal F},\reals)$ is the pointwise partial order,  
${\cal B}(X,{\cal F},\reals)$ is archimedean, and the 
Boolean algebra 
\[
{\cal P}(X,{\cal F},\reals) :=\{p\in{\cal B}(X,{\cal F},
 \reals)\mid p(x)\in\{0,1\},\forall x\in X\}
\] 
of projections in ${\cal E}(X,{\cal F},\reals)$ consists of 
the characteristic set functions $\chi\sb{M}$ of sets 
$M\in{\cal F}$.
\end{example}

Recall that a partially ordered abelian group $G$ is said to be 
\emph{lattice ordered}, or for short, is an \emph{$\ell$-group}, 
iff every pair of elements $g,h\in G$ has an infimum $g\wedge
\sb{G} h$ and a supremum $g\vee\sb{G} h$ in the partially 
ordered set $G$. The additive partially ordered abelian 
groups $C(X,\reals)$ and ${\cal B}(X,{\cal F},\reals)$ in 
Examples \ref{ex:CofX} and \ref{ex:sigmafield} are $\ell$-groups 
with pointwise minimum and maximum as the infimum and supremum, 
respectively. 

If $G$ has the property that, for every $a,b,c,d\in G$ with 
$a,b\leq c,d$ (i.e., $a\leq c,a\leq d,b\leq c,$ and $b\leq d$), 
there exists $t\in G$ with $a,b\leq t\leq c,d$, then $G$ has the 
\emph{Riesz interpolation property}, or for short, $G$ is an 
\emph{interpolation group} \cite[Chapter 2]{Good}. If $G$ 
is an $\ell$-group, then it is an interpolation group. (Just 
take $t$ to be any element between $a\vee\sb{G}b$ and 
$c\wedge\sb{G}d$.) Thus, the directed groups $C(X,\reals)$ and 
${\cal B}(X,{\cal F})$ are interpolation groups.

The so-called \emph{MV-algebras}, which play an important role in 
the analysis of many-valued logics \cite{CCC57,CCC58} and in the 
classification of AF C$\sp{\ast}$-algebras \cite{Mund}, can be  
characterized as the effect algebras that are realized as unit 
intervals in abelian $\ell$-groups with order units. Thus, the unit 
intervals $E(X,\reals)$ and ${\cal E}(X,{\cal F},\reals)$ in 
Examples \ref{ex:CofX} and \ref{ex:sigmafield} are MV-algebras. 
Not every MV-algebra can be realized as the unit interval in a 
c-ring, but the author does not know a perspicuous 
characterization of those that can.

In the theory of operator algebras, there are well-known 
connections between commutativity and lattice structure. For 
instance, by a theorem of S. Sherman \cite{Sherm}, a unital 
C$\sp{\ast}$-algebra $A$ (Example \ref{ex:CStar}) is commutative 
iff the directed group $G$ of self-adjoint elements in $A$ is an 
$\ell$-group. On the other hand, by a result of R. Kadison \cite{Kad}, 
if $A$ is a von Neumann algebra, then $A$ is a factor iff the 
directed group $G$ is an antilattice (i.e., only pairs of comparable 
elements can have an infimum or a supremum in $G$). Under suitable 
hypotheses (borrowed from the theory of operator algebras) similar 
results can be obtained for groups with compression bases \cite{FoPu};
hence for e-rings. If $P$ is a Boolean algebra, then it is a lattice, 
so Corollary \ref{cor:PBoo} already furnishes a hint of the 
commutativity-lattice connection for e-rings; further 
evidence is provided by Theorems \ref{th:ellBoo}, \ref{th:ellgroup}, 
and \ref{th:E=P} below. 

\begin{theorem} \label{th:ellBoo}
Suppose that $G$ is an $\ell$-group {\rm (}or more generally, an 
interpolation group {\rm \cite{Good})}. Then, {\rm (i)} $G\subseteq 
C(P)$ and {\rm (ii)} $P$ is a Boolean algebra.
\end{theorem}

\begin{proof} Assume that $G$ is an interpolation group.

(i) Let $0\leq g\in G$ and $p\in P$. As $G$ is directed, it will be 
sufficient to prove that $gCp$. By Lemma \ref{lm:orderunit} (i), there 
exists $n\in\Nat$ such that $0\leq g\leq n\cdot 1$, whence 
$0\leq g\leq np+n(1-p)$. As $G$ is an interpolation group, there 
exist $x,y\in G$ with $0\leq x\leq np$, $0\leq y\leq n(1-p)$, and 
$g=x+y$ (see \cite[Proposition 2.1 (b)]{Good}.  Thus, by Lemma 
\ref{lm:M} (iv), $x=xp=px$ and $y=y(1-p)=(1-p)y$, and it follows 
that $yp=py=0$ and $gp=pg=x$.

(ii) Follows from (i) and Corollary \ref{cor:PBoo}.  
\end{proof}

The c-ring in Example \ref{ex:sigmafield} satisfies the conditions 
in the following theorem. Of course, the c-ring in Example 
\ref{ex:CofX} satisfies condition (i), and it satisfies condition 
(ii) if $X$ is basically disconnected (i.e., the closure of every 
open F$\sb{\sigma}$ subset of $X$ remains open). 

\begin{theorem} \label{th:ellgroup}
Suppose that {\rm (i)} $G\subseteq C(P)$, and {\rm (ii)} for every 
$g\in G$, there exists $p\in P$ with $(1-p)g\leq 0\leq pg$. Then $G$ 
is an $\ell$-group.
\end{theorem}

\begin{proof}
The proof is adapted from \cite[Proposition 8.9]{Good}. Let $g,h\in G$. 
It will be sufficient to prove that the supremum  $g\vee\sb{G}h$ exists 
in $G$. By hypothesis, there exists $p\in P$ such that $(1-p)(g-h)\leq 0
\leq p(g-h)$. Put 
\[
s :=pg+(1-p)h=p(g-h)+h=(1-p)(h-g)+g.
\]
Evidently, $h,g\leq s$. Furthermore, if $k\in G$ with $g,h\leq k$, 
then by Lemma \ref{lm:M} (i), $pg\leq pk$ and $(1-p)h\leq(1-p)k$, 
whence $s\leq pk+(1-p)k=k$. Therefore, $s=g\vee\sb{G}h$.
\end{proof}

A subset $A\subseteq P$ is said to be \emph{orthogonal} iff, 
for all $a,b\in A$, $a\not=b\Rightarrow ab=0$. If $A$ is an 
orthogonal subset of $P$, then by Corollary \ref{cor:pqinP} 
and induction on $n$, the sum $p :=a\sb{1}+a\sb{2}+\cdots+
a\sb{n}$ of finitely many distinct elements $a\sb{1},a\sb{2},
...,a\sb{n}\in A$ belongs to $P$ and coincides with the 
supremum $p :=a\sb{1}\vee a\sb{2}\vee\cdots\vee a\sb{n}$ 
both in $P$ and in $E$.

\begin{theorem} \label{th:E=P}
The following conditions are mutually equivalent:
\begin{enumerate}
\item $P$ is a Boolean algebra and $P$ generates the group $G$.
\item $G\subseteq C(G)$ and, for each $g\in G$, there is a 
 finite orthogonal set $A\subseteq P$ such that $g$ is a 
 linear combination with integer coefficients of the elements of 
 $A$.
\item $G$ is an $\ell$-group and $E=P$.
\item $G$ is an interpolation group and $1$ is a minimal order 
 unit in $G$.
\item $E$ is a Boolean algebra with $e\mapsto 1-e$ as the 
 Boolean complementation mapping.
\item $E=P$.
\end{enumerate}
\end{theorem}

\begin{proof}
(i) $\Rightarrow$ (ii). Assume (i). By Corollary \ref{cor:PBoo}, 
$P\subseteq C(P)$ and, since $P$ generates $G$, it follows that 
$G\subseteq C(G)$.  Let $g\in G$. Then there are projections 
$p\sb{i}\in P$, $1\leq i\leq n$, and integer coefficients 
$c\sb{i}$ such that $g=\sum\sb{i=1}\sp{n}c\sb{i}p\sb{i}$.
Let $B$ be the sub-Boolean algebra of $P$ generated by $p\sb{i}$,  
$1\leq i\leq n$. Since $B$ is a finitely generated Boolean 
algebra, it is finite. Let $A$ be the set of atoms (minimal 
nonzero elements) in $B$. Then, if $a,b\in A$ with $a\not=b$, we 
have $ab=ba=a\wedge b=0$ (Theorem \ref{th:pqinP}), so $A$ is a 
finite orthogonal subset of $P$. Also, each element in $B$, and 
in particular each $p\sb{i}$, can be written as a sum of certain 
of the projections in $A$.  Thus, by gathering terms, we can 
write $g=\sum\sb{i=1}\sp{n}c\sb{i}p\sb{i}=\sum\sb{a\in A}k\sb{a}a$ 
with integer coefficients $k\sb{a}$ for all $a\in A$.

(ii) $\Rightarrow$ (iii). Assume (ii). Then $G\subseteq C(G)
\subseteq C(P)$. Let $g\in G$, and let $A$ be a finite orthogonal 
subset of $P$ such that $g=\sum\sb{a\in A}k\sb{a}a$ for integer 
coefficients $k\sb{a}$. Define $A\sb{+} :=\{a\in A\mid k\sb{a}>0\}$, 
$A\sb{-} :=\{a\in A\mid k\sb{a}<0\}$, and $p :=\sum\sb{a\in A\sb{+}}a$. 
Then $p\in P$, $a\in A\sb{+}\Rightarrow pa=k\sb{a}a$, and $a\in A\sb{-}
\Rightarrow pa=0$. Thus, $pg=\sum\sb{a\in A\sb{+}}k\sb{a}a\geq 0$ and 
$(1-p)g=g-pg=\sum\sb{a\in A\sb{-}}k\sb{a}a\leq 0$; hence $G$ is an 
$\ell$-group by Theorem \ref{th:ellgroup}. Now suppose $g\in E$. 
Then, if $a\in A\sb{-}$, we have $0\leq ga=ag$, whence $0\leq ga=
k\sb{a}a\leq 0$, so $k\sb{a}a=0$. Consequently, $g=\sum\sb{a\in 
A\sb{+}}k\sb{a}a$. Also, if $a\in A\sb{+}$, we have $a\leq k\sb{a}a
=ga\leq a$ (Lemma \ref{lm:FF} (i)), so $k\sb{a}a=a$. Consequently, 
$g=\sum\sb{a\in A\sb{+}}k\sb{a}a=\sum\sb{a\in A\sb{+}}a=p\in P$. 
Therefore, $E=P$. 

(iii) $\Rightarrow$ (iv).  Assume (iii). Then $G$ is an interpolation 
group. Suppose $p$ is an order unit in $G$ and $p\leq 1$. Then 
$0\leq p$ and there exists $n\in\Nat$ such that $0\leq 1-p\leq np$.
As $0\leq p\leq 1$, we have $p\in E=P$; hence by Lemma \ref{lm:M} (iv), 
$1-p=(1-p)p=0$, i.e., $p=1$.  Thus $1$ is a minimal order unit in $G$.

(iv) $\Rightarrow$ (v). Assume (iv). By Theorem \ref{th:ellBoo}, 
$G\subseteq C(P)$ and $P$ is a Boolean algebra with $p\mapsto 1-p$ 
as the Boolean complementation. It will be enough to show that 
$E=P$. Thus, let $e\in E$ and suppose that $d\in E$ with $d\leq 
e,1-e$. Then $1-e\leq 1-d$ and $e\leq 1-d$. Adding the last two 
inequalities, we find that $1\leq 2(1-d)$. Thus, if $g\in G$, there 
exists $n\in\Nat$ such that $g\leq n\cdot 1\leq 2n(1-d)$, and it 
follows that $1-d$ is an order unit in $G$. But $1-d\leq 1$; hence,  
by hypothesis, $1-d=1$, i.e., $d=0$. Consequently, $e\in P$ by 
Theorem \ref{th:sharp}, and we conclude that $E=P$.

(v) $\Rightarrow$ (vi).  Assume (v) and let $e\in E$. If $d\in E$ 
with $d\leq e,1-e$, then since $1-e$ is the Boolean complement of 
$e$ in the Boolean algebra $E$, it follows that $d=0$; hence 
$e\in P$ by Theorem \ref{th:sharp}.  Consequently, $E=P$.

(vi) $\Rightarrow$ (i). Assume (vi) and let $p,q\in P=E$. Since $E$ 
generates the group $G$, it will be sufficient by Corollary \ref
{cor:PBoo} to prove that $pCq$.  Let $d :=pqp$.  By Lemma \ref{lm:FF} 
(ii), $d\in E=P$, and it is clear that $d=dp=pd$, whence $d\leq p$. 
Thus, $dqd=dpqpd=d\sp{3}=d$, and it follows that $d(1-q)d=0$. 
Therefore, $d(1-q)=0$, i.e., $d=dq$, so $d\leq q$. By Corollary 
\ref{cor:q-pinP}, $p\sb{1} :=p-d\in P$ and $q\sb{1} :=q-d\in P$. 
Also, $pq\sb{1}p=pqp-pdp=d-d=0$, so $pq\sb{1}=0$, $p=d+p\sb{1}$, 
$q=d+q\sb{1}$, and by Theorem \ref{th:p+qinP} $d+p\sb{1}+q\sb{1}
=p+q\sb{1}\in P$.  Consequently, $p$ and $q$ are Mackey compatible, 
so $pCq$ by Lemma \ref{lm:pCq}.  
\end{proof}

If the e-ring $(R,E)$ satisfies any, hence all, of the conditions 
(i)--(v) in Theorem \ref{th:E=P}, then $G\subseteq C(G)$ by condition 
(ii), and we can drop down to the c-ring $(G,E)$, which of course 
will continue to satisfy conditions (i)--(v).  

\begin{definition}
\textnormal{A \emph{b-ring} is a c-ring $(G,E)$ satisfying any, 
hence all, of the conditions (i)--(v) in Theorem \ref{th:E=P}.}
\end{definition}

The b-ring in the following example is a modification of Example 
\ref{ex:sigmafield} in which the totally ordered field $\reals$ is 
replaced by the totally ordered ring $\integers$ of integers and 
the $\sigma$-field ${\cal F}$ is replaced by any field of sets. 

\begin{example} \label{ex:field}
Let ${\cal F}$ be a field of subsets of a nonempty set $X$, define 
${\cal B}(X,{\cal F},\integers)$ to be the ring under pointwise 
operations of all bounded functions $f\colon X\to\integers$ such that 
$f\sp{-1}(z)\in{\cal F}$ for all $z\in\integers$, and let 
\[
 {\cal E}(X,{\cal F},\integers) :=\{e\in{\cal B}(X,{\cal F},
 \integers)\mid e(x)\in\{0,1\},\forall x\in X\}.
\]
Then $({\cal B}(X,{\cal F},\integers),{\cal E}(X,{\cal F},\integers))$ 
is a b-ring, the partial order on ${\cal B}(X,{\cal F},\integers)$ 
is the pointwise partial order, and ${\cal B}(X,{\cal F},\integers)$ 
is archimedean. Thus the effects in ${\cal E}(X,{\cal F},\integers)$,  
which coincide with the projections for the b-ring $({\cal B}(X,{\cal F},
\integers),\\{\cal E}(X,{\cal F},\integers))$, are the characteristic 
set functions $\chi\sb{M}$ of sets $M\in{\cal F}$.
\end{example}

Under set-inclusion, a field ${\cal F}$ of subsets of a nonempty 
set $X$ is a Boolean algebra, and in Example \ref{ex:field}, the 
Boolean algebra ${\cal E}(X,{\cal F},\integers)$ is isomorphic to 
${\cal F}$. By the Stone representation theorem, every Boolean 
algebra $B$ is isomorphic to the field ${\cal F}$ of compact open 
subsets of a compact Hausdorff totally-disconnected space $X$; 
hence, \emph{every Boolean algebra can be realized as the Boolean 
algebra of projections in a b-ring.}

The functions $f\colon X\to\integers$ in Example \ref{ex:field} can 
be regarded as ``signed multisets" by thinking of $f(x)$ as the 
``signed multiplicity of $x$ in $f$."  In \cite{TH}, T. Hailperin 
suggests that, in contemporary algebraic terms, the true realization of 
Boole's original ideas is not what is now called a Boolean algebra, 
but rather it is an algebra of signed multisets forming a commutative 
ring with unity and with no nonzero additive or multiplicative 
nilpotents. Our b-ring $({\cal B}(X,{\cal F},\integers),{\cal E}
(X,{\cal F},\integers))$ is precisely such an algebra, and the 
``b" in ``b-ring" is meant to suggest this Boolean connection.

\begin{theorem} \label{th:Booext}
Let $(G,E)$ and $(H,F)$ be b-rings and let $\phi\colon E\to F$ be a 
Boolean homomorphism from the Boolean algebra $E$ to the Boolean 
algebra $F$. Then $\phi$ admits a unique extension to a group 
homomorphism $\Phi\colon G\to H$ of the additive group $G$ into 
the additive group $H$.  Moreover, $\Phi\colon G\to H$ is an 
order-preserving ring homomorphism with $\Phi(1)=1$.
\end{theorem}

\begin{proof}
The Boolean homomorphism $\phi\colon E\to F$ preserves $0$, $1$, 
finite infima, and finite suprema. For $p,q\in E=P$, we have 
$p\wedge q=pq$; hence $\phi(pq)=\phi(p\wedge q)=\phi(p)\wedge\phi(q)
=\phi(p)\phi(q)$, i.e., $\phi$ preserves products of projections. 
Also, if $p+q\in E$, then $p\vee q=p+q$; hence $\phi(p+q)=\phi(p\vee q)
=\phi(p)\vee\phi(q)=\phi(p)+\phi(q)$, i.e., $\phi\colon E\to H$ 
preserves existing sums in $E$. Since $G$ is an interpolation group, 
a theorem of S. Pulmannov\'{a} \cite{Pu99} implies that $\phi$ admits 
a unique extension to a group homomorphism $\Phi\colon G\to H$. As 
$\phi(E)\subseteq F$, it follows that $\Phi(E\sp{+})\subseteq F\sp{+}$, 
whence $\Phi$ is order preserving.  Every element in $G$ is a finite 
linear combination of projections with integer coefficients, and since 
$\Phi$ preserves products of projections, it follows that $\Phi$ 
preserves products. Obviously, $\Phi(1)=\phi(1)=1$.
\end{proof}

The following is the fundamental structure theorem for b-rings.

\begin{theorem} \label{th:bring}
Let $(G,E)$ be a b-ring, let $X$ be the Stone space of the Boolean  
algebra $E$, and let ${\cal F}$ be the field of compact open subsets 
of $X$. Then there is an order and ring isomorphism $\Phi\colon G
\to{\cal B}(X,{\cal F},\integers)$ such that the restriction 
$\phi$ of $\Phi$ to $E$ is a Boolean isomorphism of $E$ onto 
${\cal E}(X,{\cal F},\integers)$. 
\end{theorem}

\begin{proof}
The projections in ${\cal E}(X,{\cal F},\integers)$ are characteristic 
set functions $\chi\sb{K}$ of compact open subsets $K$ of $X$; hence 
by Stone's representation theorem, there is a Boolean isomorphism 
$\phi\colon E\to{\cal E}(X,{\cal F},\integers)$. By Theorem 
\ref{th:Booext}, $\phi$ can be extended to an order-preserving 
ring homomorphism $\Phi\colon G\to{\cal B}(X,{\cal F},\integers)$ 
and $\phi\sp{-1}\colon{\cal E}(X,{\cal F},\integers)\to E$ can be 
extended to an order-preserving ring homomorphism $\Psi\colon{\cal B}
(X,{\cal F},\integers)\to G$.  The ring endomorphism $\Psi\circ\Phi
\colon G\to G$ is the identity on $E$, and $E$ generates $G$; hence 
$\Psi\circ\Phi\colon G\to G$ is the identity on $G$. Likewise 
$\Phi\circ\Psi\colon$ is the identity on ${\cal B}(X,{\cal F},
\integers)$, so $\Phi$ is an order-preserving ring isomorphism 
with $\Psi=\Phi\sp{-1}$.
\end{proof}

\end{document}